\documentclass[10pt]{article}

\usepackage{amsmath}  
\usepackage{graphicx}
\usepackage{caption}
\usepackage{subfig}
\usepackage{color}
\usepackage[margin=1in]{geometry}
\usepackage{multirow}
\usepackage{tabularx}
\usepackage{cite}
\usepackage[font=footnotesize,labelfont=bf]{caption}
\usepackage{float}
\usepackage{amssymb}
\usepackage{amsmath}

\title{\bf{There is Plenty of Room for THz Tunneling Electron Devices Beyond the Transit Time Limit}}
\author{{Matteo Villani}$^{1}$, {Simone Clochiatti}$ ^{2}$, {Werner Prost}$^{2}$,  {Nils Weimann}$^{2}$, {Xavier Oriols}$^{1}$}
\date{\normalsize{1: Department of Electronic Engineering, Universitat Aut\`onoma de Barcelona, Campus de la UAB, 08193 Bellaterra, Barcelona, Spain \\ 2: Components for High Frequency Electronics (BHE), University of Duisburg-Essen, 47057 Duisburg, Germany}}
\begin{document}
\maketitle
%\makeauthor
\begin{abstract}
	
	The traditional transmission coefficient present in the original Landauer formulation, which is valid for quasi-static scenarios with working frequencies below the inverse of the electron transit time, is substituted by a novel time-dependent displacement current coefficient valid for frequencies above this limit. Our model captures in a simple way the displacement current component of the total current, which at frequencies  larger than the inverse of the electron transit time can be more relevant than the particle component. The proposed model is applied to compute the response of a resonant tunneling diode from 10$\,$GHz up to 5$\,$THz. We show that tunneling electron devices are intrinsically nonlinear at such high frequencies, even under small-signal conditions, due to memory effects related to the displacement current. We show that these intrinsic nonlinearities (anharmonicities) represent an advantage, rather than a drawback, as they open the path for tunneling devices in many THz applications, and avoid further device downscaling.
\end{abstract}

\paragraph{Corresponding author:}$\!\!$xavier.oriols@uab.es

\section{Introduction}
For high enough frequencies, it is said that electron devices behave as low pass filters. This behavior starts beyond the so-called transit time limit. %, when the inverse of the maximum working frequency becomes comparable to the electron transit time.
Microscopically, this limit is reached when electrons crossing the device are not affected by a static potential, but a time-dependent one. The usual strategy to design electron devices working at high frequencies is reducing their size to obtain smaller transit times.

We argue in this work that there is plenty of room for quantum devices to reach higher working frequencies (without further miniaturization) by designing them to work beyond the transit time limit. Beyond such limit, the output signal is not able to exactly follow the input signal, this produces nonlinearities that can be an advantage (rather than a drawback) for many applications like frequency multipliers, rectifiers, oscillators, and in general any signal modulator. Indeed, there are already prototypes of resonant tunneling diodes (RTD) as THz devices working at output frequencies close \cite{Brown_RTD, Arzi,Wasige,Nagatsuma} or even beyond \cite{Feiginov1,Asada1} the mentioned transit time limit. 

The THz nonlinearity (anharmonicity) that will be shown in this work appears in any ballistic device, as it needs to avoid the randomness due to collisions in the active region. Tunneling devices are chosen in this work, since they offer enriched quantum coherent electron dynamics by engineering the active region under AC conditions. We stress that the THz nonlinearity discussed in this work is unrelated to any nonlinearity present in the static characteristics of RTD devices.

The Landauer formula \cite{Landauer1} based on the transmission coefficient has been a very simple and powerful tool to predict the response of tunneling devices below the transit time limit. Beyond the transit time limit, the displacement current component is known to become relevant over the particle one. This contribution has been modeled previously using a master equation \cite{Feiginov2001} or a scattering matrix \cite{reviewButtiker,Landauer2} to deal with electron transitions between different parts of the tunneling device. In this work the electron is evolving in the whole tunneling structure, while keeping its full-coherence \cite{IWMTS_2020, Oriols_insp, Oriols}.

A simple, general and accurate model is presented for the computation of the total current in quantum coherent devices that just substitutes the traditional static transmission coefficient of the Landauer formula with the time-dependent displacement current coefficient. The model presented here, by construction valid beyond the transit time limit, shall spread the message of the title of this paper, opening intuition for quantum engineers to use the displacement current for new THz applications in electron tunneling devices.

\section{Displacement current coefficient}
We consider in this work a two-terminal device with an active region of length $L=b-a$ defined as the space between source and drain contacts ($a<x<b$). For  DC computations, the \textit{transmission coefficient} $T$ of the $i-$th electron injected from the left ($x<a$), at time $t_i$, is expressed by,
\vspace{-0.1cm}
\begin{equation}
T=\int_{b}^{\infty}|\psi_i(x,t_i+\tau')|^2\;dx=\int_{t_i}^{t_i+\tau'}J_i(b,t) dt
\label{eq2}
\end{equation}

where $|\psi_i(x,t)|^2$ is the quantum probability distribution of a wave-packet, $J_i(b,t)$ is the associated current density as defined in \cite{cohen} calculated at the drain contact surface $x=b$. Finally $\tau'$ is a time interval large enough to ensure that the $i$-th electron has completely crossed the active region. Notice that we do not need to anticipate which is the exact electron transit time $\tau$ (which can be larger than the tunneling time \cite{Tunneling1,Tunneling3}) in Eq. \ref{eq2} because the same value $T$ will be obtained with whatever $\tau'$ satisfying $\tau' >\tau $ (for example $\tau' \to \infty$). To capture the dynamic behavior of electrons in THz scenarios, a Bohmian approach is here used\cite{Bohm,Oriols1,Damiano} where each electron is described by a defined trajectory $x_i(t)$. The total current measured at the contacts of a two-terminal device is described by the Ramo-Shockley theorem \cite{Ramo,Sho} as
\vspace{-0.1cm}
\begin{equation}
I(t)=\frac{q }{L} \sum_{i=1}^{N(t)} v_i(x_i(t),t)
\label{eq3}
\end{equation}
where $q$ is the electron charge and $N(t)$ is the number of electrons inside the device. The velocity $v_i(x,t)={J_i(x,t)}/{|\psi_i(x,t)|^2}$ of each electron $x_i(t)$ is calculated \cite{Bohm,Oriols1} from the wavepacket $\psi_i(x,t)$. We can see that \eqref{eq3} takes into account the displacement current because the movement of any $i-$th electron with $x_i(t)\in[a,b]$  contributes to the total current at time $t$  (by generating a time-dependent electrical field everywhere \cite{Ramo}).

The parameter $N(t)$ does not need an explicit calculation because the number of electrons inside the active region given by the sum in \eqref{eq3} can be transformed into an integral over the probability of presence in the active region at time $t$ for every $i$-th electron entering at time $t_i<t$. In other words: $\sum_{i=1}^{N(t)} \rightarrow \int_{t-\tau'}^{t} \gamma \, dt_i \int_{a}^{b} |\psi_{i}(x,t)|^2\; dx$ where $\gamma$ is a parameter proportional to the lateral area of the device \cite{Oriols_insp}. Ignoring the energy integrals in the evaluation of the current (that would only obscure the conceptual discussion of this paper), the total current in \eqref{eq3} can be written as $I^f(t)= q D^{f}(t)$ with the \textit{displacement current coefficient} given by
\begin{equation}
D^{f}(t) \equiv \frac{\gamma}{L} \,  \int^{t}_{t-\tau'} dt_i \int^{b}_{a} dx J_{i}^f(x,t)=\int^{t}_{t-\tau'}dt_i G^f(t,t_i)
\label{eq7}
\end{equation}
The exact evaluation of $D^{f}(t)$ in Eq. (3) is as follows:
\begin{itemize} \item The $i$-th electron is represented by an initial Gaussian wavepacket $\psi_i(x,t_i)$ at time $t_i$ located outside of the active region. 
	\item The time-dependent Schr\"odinger equation with a potential profile oscillating at frequency $f$ is solved to get $\psi_i(x,t)$ and $J_{i}^f(x,t)$ at all times $[t_i,t_i+\tau']$. 
	\item The contribution of the $i$-th electron to $D^f(t)$ evaluated from \eqref{eq7} is proportional to the amount of current density $J_{i}^f(x,t)$ inside the active region at time $t$. 
\end{itemize}
The above three steps are repeated for the train of electrons with identical properties but different injecting times in the range $t_i\in[\, t\, ,\,t-\tau']$. The injection time $t_i$ is considered a continuous variable with a numerical discretization for practical computation.
The $G^f(t,t_i)$  captures the "memory" effects and it quantifies the contribution of the $i$-th electron entering at $t_i$ on the total current at time $t$ for the external input frequency $f$.
We specify again that no physical value of the exact electron transit time $\tau$ is needed in the computation of Eq. \ref{eq7}. In fact, the same value $D^{f}(t)$ will be obtained for any value of $\tau'$  as far as it satisfies $\tau' >\tau $, where $\tau$ is the electron transit time. A reasonable estimation of the physical transit time of the $i$-th electron can be obtained by the usual expression $\tau=\int^{t_i+\tau'}_{t_i} dt \int^{b}_{a} dx |\psi_{i}(x,t)|^2$ with $\tau' >\tau$.

\section{Small-signal and THz nonlinearity in time}

\begin{figure}[b]
	\centering
	\includegraphics[scale=0.1]{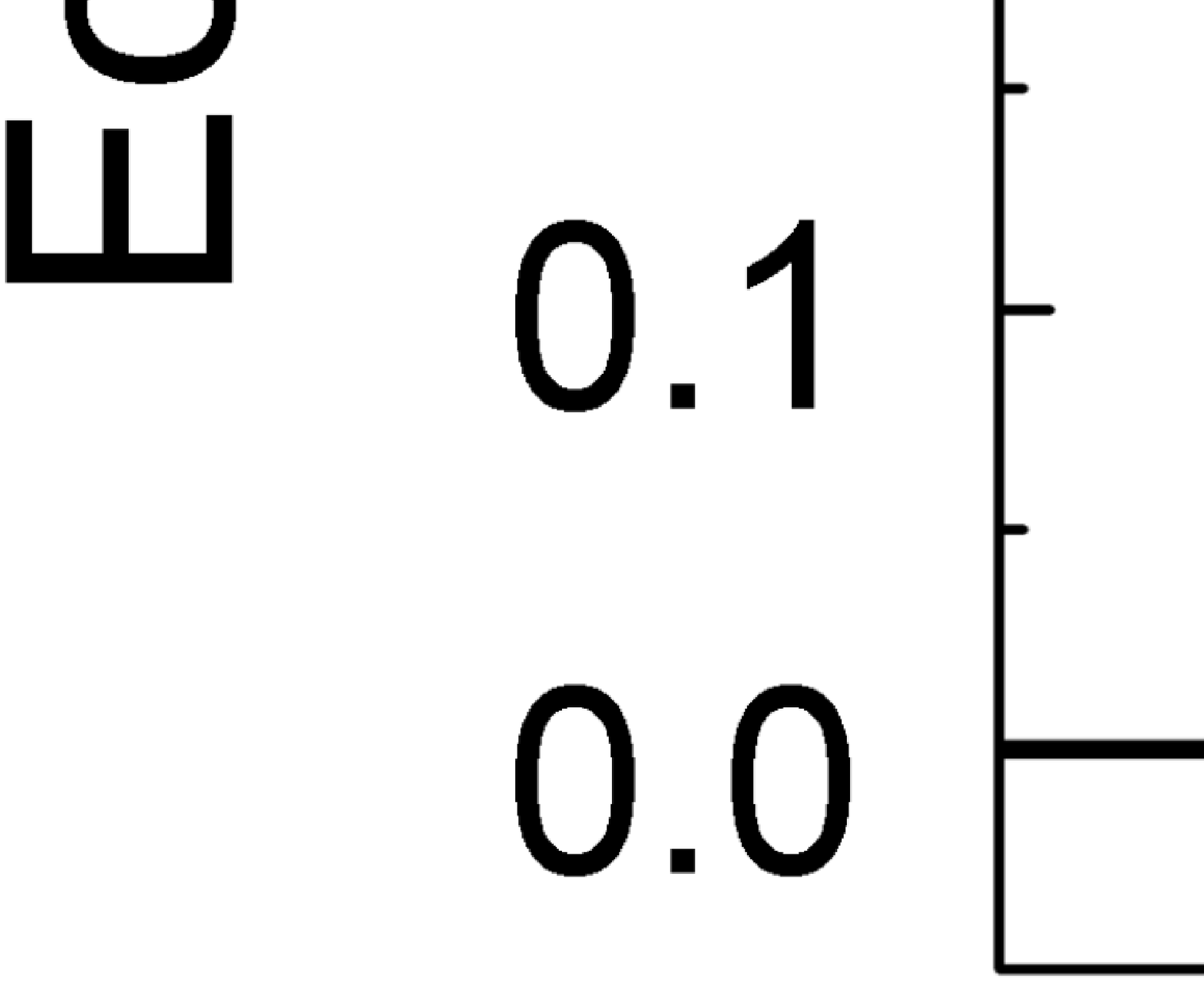}
	\caption{(a) Band structure of GaAs/AlGaAs RTD device with 1.7 nm for both, barrier thickness and well width. (b) DC I-V characteristic of the RTD device.}
	\label{fig0}
\end{figure}

In a quasi-static discussions of device performance, only the input signal amplitude becomes relevant to determine a linear behavior. We anticipate that linearity in our high frequency discussion will depend also on the input frequency $f$. The same small input signal amplitude that gives a linear behavior at a given low frequency can provide a non-linear one at higher frequency. In this section, we compute the time-dependent total current $I(t)$ for a RTD device whose potential profile $E_c(x)$ at zero bias is plotted in Fig. \ref{fig0} together with its typical \cite{Brown_RTD} DC characteristic with a current peak at $0.75 \, V$. 

A signal  $V_{in}(t)=V_0 \, cos (2 \pi  f \, t+ \phi)$ is applied at the drain (with the source contact grounded) with $V_0=0.01 \, V$, which is a small enough value to have a linear I-V relationship. The time-dependent potential profile given by $E_c(x,t)=E_c(x)-x \cdot q \, \frac{V_0}{L} \, cos (2 \pi f \, t+ \phi) $ is considered, with $\phi$ the phase of the $V_{in}$ signal. A non self-consistent potential profile is simulated, to avoid complications that will not change the main result of this work.
It is worth underlining that the time dependent Schr\"odinger equation giving the evolution of the wavepackets $\psi_i(x,t)$ and $J_{i}^f(x,t)$ to compute $D^f(t)$ in Eq. \ref{eq7} is driven by the time dependent Hamiltonian $H(x,t)=-\frac{\hbar^2}{2m} \frac{\partial^2}{\partial x^2}+E_c(x,t)$ which has no energy eigenstates because the electron total energy can vary locally during the electron transit along the active region. Certainly, such electron dynamics cannot be reached with the energy eigenstates linked to the time-independent Hamiltonians used in the Landauer model.

\begin{figure}[h]
	\vspace{-0.3cm}
	\centering
	\includegraphics[scale=0.4]{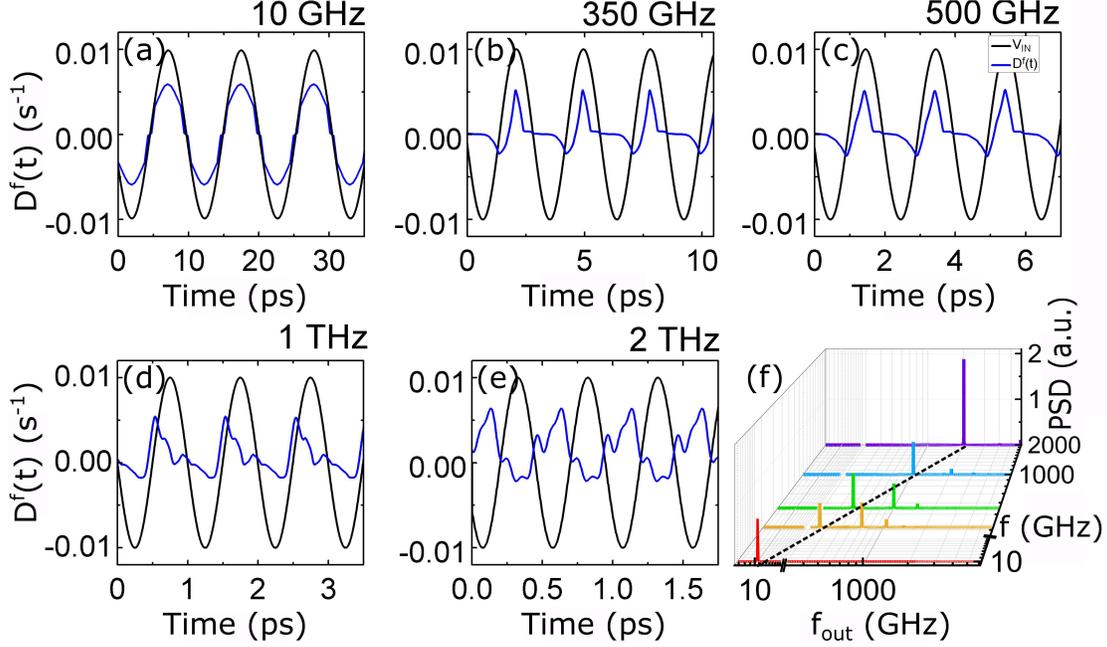}
	\caption{(a)-(e) The displacement current coefficient D$^{f}$(t), proportional to the output current I(t), is plotted (in blue) as a function of time for different input frequencies f, with the input small-signal AC voltage V$_{in}$(t) (dashed black in arbitrary units). (f) Power spectral density (PSD) of the output currents of (a)-(e) as a function of the output frequency f$_{out}$ confirming the harmonic generation of a nonlinear response.}
	\label{fig5}
\end{figure}

The result in Fig. \ref{fig5} (a) shows the $D^f(t)$ coefficient for low input frequency of 10 GHz, here the response is linear because $f<<1/\tau$.
For Fig. \ref{fig5} (b)-(e), the nonlinear response can be seen in time domain. Frequency multiplication is observed in Fig. \ref{fig5} (b)-(d) and confirmed in Fig. \ref{fig5} (f), where the Power Spectral Density (PSD) of (a)-(d) is shown for every input frequency $f$. We recall that small signal conditions around zero bias are applied so that the nonlinearities are not due to the DC characteristics of the RTD exhibiting strong nonlinearity in Fig. \ref{fig0}. The nonlinearity originates from the fact that the $i$-th electron entering inside the device at time $t_i \approx t-\tau$ and another $i'$-th electron entering at time $t_{i'}\approx t$ are affected by different potential profiles during their evolution. This produces different contributions to the total current $I(t)$ in Eq. \ref{eq3}. In other words, the quantum coherence of electron and displacement current ensure that the current $I(t)$ at time $t$ is influenced not only by the potential profile $E_c(x,t)$, but also by $E_c(x,t-\tau)$. This is the ``memory'' effect leading to nonlinearity for high input frequency, even when a very small input signal amplitude is used.

\section{Small-signal THz nonlinearity in frequency}

We investigate now the same small-signal nonlinear RTD response in the frequency domain. 
When the RTD is excited by a steady-state sinusoidal signal with frequency $f$, the output would be a sinusoidal signal with the same output frequency $f_{out}=f$ but different amplitude and phase. These would be described by a small signal conductance $Y_{11}(f)$ given by $I^{f}_R(t)+j \cdot I^{f}_I(t)=V_0 \, Y_{11} \cdot  \, e^{j 2 \pi f\, t}$, with $I^{f}_R(t)$ and $I^{f}_I(t)$ the real and imaginary components of the response to the small-signal $V_0 \cdot \, e^{j 2 \pi f\, t}$. See \cite{Y_parameter} for further details. We now explicitly rewrite the real part of the input signal $V_0 \, cos(2\pi f\, t)$ and the imaginary part $V_0 \, cos(2\pi f\, t-\pi/2)$. Then, using $I^f(t)= q D^{f}(t)$, one gets:
\begin{equation}
Y_{11}=\frac{I^{f}_R(0)}{V_0}+j \frac{I^{f}_I(0)}{V_0}=q\bigg(\frac{D^f_{0}}{V_0}+j\frac{D^f_{-\pi/2}}{V_0} \bigg)
\label{eq11}
\end{equation}
where the subindex $\phi=0$ and $\phi=-\pi/2$ indicates the two phases (cosine and sine respectively) corresponding to the real and imaginary parts of the input signal. To simplify the notation, if $t=0$, the  $t$ in $D^f(t)$ and $G^f(t,t_i)$ from Eq. \ref{eq7} will be omitted. 

\begin{figure}[h]
	\centering
	\includegraphics[scale=0.2]{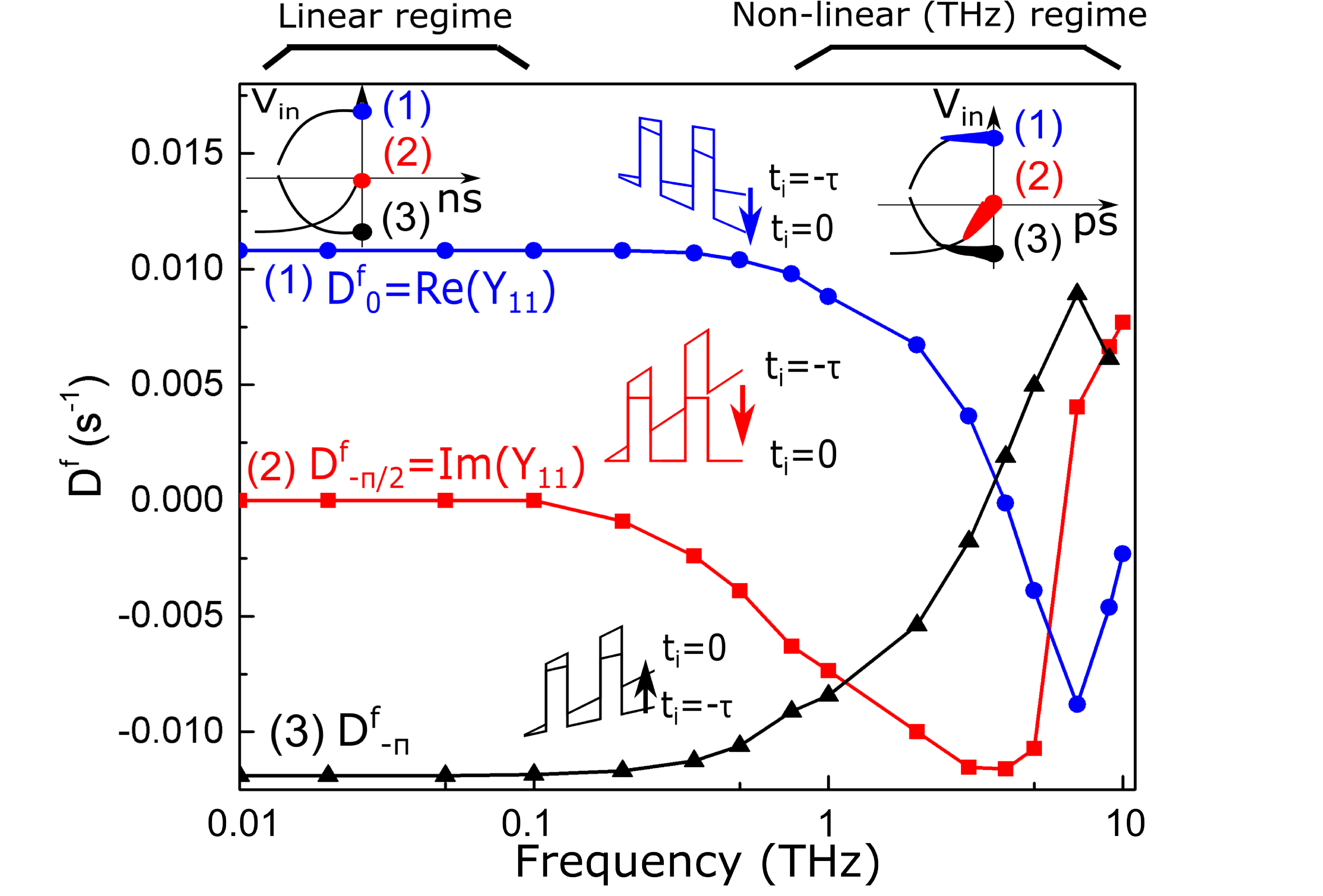}
	\caption{The displacement current coefficient D$^f$ of the RTD with an input signal V$_{in}$(t) for different frequencies $f$ estimated at phases: $\phi$=$0$ (blue line), $\phi$=$-\pi/2$ (red line) and $\phi$=$-\pi$ (black line). For each phase, the evolution of the potential profile E$_c(x,t)$ of the RTD is shown. The left and right top insets show in black the input signal \emph{before} the present time $t$ for each of the three phases, and in colors the relevant times involved in the evaluation of D$^f(t)$ for low and high frequencies respectively. Linear regime (top-left inset) is achieved when the signal value V$_{in}$(t$_i$) at t$_i\approx \,\, $t$-\tau$ is equal to the one at the present time t, whereas a nonlinear regime (top-right) is reached when V$_{in}$(t$-\tau$)$\neq$ V$_{in}$(t).}
	\label{fig4}
\end{figure}

In Fig. \ref{fig4}, the $Y_{11}$ parameter defined in Eq. \ref{eq11} is plotted. The behavior between $50\,\,$GHz and $300\,\,$GHz of the real (blue) and imaginary (red) components of $Y_{11}$ is similar to the response of a device with a delay due to the transit (or switching) time $\tau$ \cite{Zhen}. In fact, it can be shown that our model exactly reproduces  the Landauer model when $f \ll 1/\tau$. In this quasi-static regime, it can be shown that $D^{f}(t)=1/L \int_{a}^{b} dx \int _{t-\tau}^{t} \gamma \, J_{i}(x,t-t_i) dt_i= \gamma \, T$.
The results in Fig. \ref{fig4} are related to the PSD shown in Fig. \ref{fig5}(f) by the dashed black line at $f_{out}=f$. With the small amplitude condition applied here, someone would assume a linear response. However, it can be proved that this is not the case simply by computing the output current related to the input signal $V_0 cos(2\pi f\; t-\pi)$, estimated with $D^f_{-\pi}$ plotted in black in Fig. \ref{fig4}. For a linear small signal device, one would expect $D^f_{-\pi}=-D^f_{0}$ because the input signals satisfy $cos(2\pi f\; t-\pi)=-cos(2\pi f\; t)$. This is true at low frequencies but at $f>200\,\,$GHz differences appear, proving nonlinearity. The frequency range $f>200\,\,$GHz is recently being explored for real RTD-based THz sources and detectors\cite{Wasige,Arzi,Nagatsuma}.
The physical and unavoidable reason of this small-signal nonlinearity at $f>200\,\,$GHz is explained by the function $G^f(t_i)$  in Eq. \ref{eq7} and the evolution of the energy potential profiles plotted in the insets of Fig. \ref{fig4}. It can be seen how electrons injected at $t_i=-\tau$  travel faster under the (blue) potentials linked to $G_0^f(-\tau)$ than under the (black) potentials linked to $G_{-\pi}^f(-\tau)$. The blue potentials accelerate electrons due to $E_c(x,-t_i)>E_c(x,0)$, while the black potentials decelerate them due to $E_c(x,-t_i)<E_c(x,0)$.
It follows that, the memory-related effects appear at smaller frequencies for $D^f_{-\pi}$ (as low as $100\,\,$GHz) than for $D^f_{0}$, as seen in Fig. \ref{fig4}.

\section{Conclusion}
In this paper, we present a model that substitutes the transmission coefficient in the Landauer formula by a new displacement current coefficient to capture realistic predictions of the behavior of tunneling devices at frequencies comparable or higher than the electron transit time (where the displacement current matters). The model is then used to show the nonlinear or anharmonic  behavior of an RTD device in small signal condition around zero bias in a frequency window from $200\,\,$GHz to $5\,\,$THz. This frequency regime is here defined as the nonlinear (THz) regime, and it remains mainly unexplored by the scientific community. We argue that, depending on the proper engineering of the device that tailors the displacement current coefficient, different THz applications can be envisioned, as frequency multipliers, rectifiers, oscillators, modulators etc.

\paragraph{Acknowledgement}
This work was supported by Spain's "Ministerio de Ciencia,
Innovacion y Universidades" under Grant No. RTI2018-097876-B-C21 (MCIU/AEI/FEDER, UE), the "Generalitat de Catalunya" and FEDER for the project 001-P-001644 (QUANTUMCAT), the "Deutsche Forschungsgemeinschaft" within the Collaborative Research Center SFB/TRR 196MARIE (Project C02), the European Union's Horizon 2020 research and innovation programme under Grant No. 881603 GrapheneCore3 and under the Marie Sk\l{}odowska-Curie Grant No. 765426 TeraApps. (Corresponding author: Matteo Villani)

\end{document}